# Numerical modeling of the multi-stage Stern–Gerlach experiment by Frisch and Segrè using co-quantum dynamics via the Schrödinger equation

Zhe He[†], Kelvin Titimbo[†], David C. Garrett[†], S. Suleyman Kahraman, and Lihong V. Wang[*]

*Caltech Optical Imaging Laboratory, Andrew and Peggy Cherng Department of Medical Engineering, Department of Electrical Engineering, California Institute of Technology, 1200 E. California Blvd., MC 138-78, Pasadena, CA 91125, USA*

† These authors contributed equally.

* Correspondence should be addressed to L.V.W. (LVW@caltech.edu).

## Abstract

We use a theory termed co-quantum dynamics (CQD) to numerically model spin flip in the multi-stage Stern–Gerlach (SG) experiment conducted by R. Frisch and E. Segrè. This experiment consists of two Stern–Gerlach apparatuses separated by an inner rotation chamber that varies the fraction of spin flip. To this day, quantum mechanical treatments inadequately predict the Frisch–Segrè experiment. Here, we account for electron-nuclear interactions according to CQD and solve the associated Schrödinger equation. Our simulation outcome agrees with the Frisch–Segrè experimental observation and supports CQD as a potential model for electron spin evolution and collapse.

## Keywords

Stern–Gerlach experiment, collapse theory, spin flip, co-quantum dynamics

## Introduction

In what is now the prototypical example of quantum measurement [1]–[3], the experiment of Stern and Gerlach in 1922 provided evidence of the quantization of spin. A decade later, Frisch and Segrè extended this experiment to include two Stern–Gerlach (SG) stages separated by an inner rotation (IR) chamber with rapidly rotating magnetic fields, resulting in partial spin flipping [4], [5]. The quantum mechanical models of the Frisch–Segrè experiment include those by E. Majorana [6] and I. I. Rabi [7]–[9]. Majorana's formula is similar to the Landau–Zener formula [10], [11], to which Rabi added the effects of electron-nuclear spin interaction. Surprisingly, these treatments inadequately explain the experimental observation.

To predict the experimental observation, the co-quantum dynamics (CQD) theory was developed by introducing the following concepts [12], [13]. CQD considers the interaction between the electron ($\vec{\mu}_e$) and nuclear ($\vec{\mu}_n$) magnetic moments through the torque-averaged instead of self-averaged magnetic field, introduces an induction term, and treats $\vec{\mu}_n$ as a non-collapsing co-quantum of the principal quantum $\vec{\mu}_e$. The strength of the electron–nuclear coupling depends on the magnitudes and relative orientations of $\vec{\mu}_e$ and $\vec{\mu}_n$ [14]. Although it is weaker than spin-orbit coupling, this coupling is important for an electron spin in the S state where the orbital angular momentum vanishes on average [15]. The electron spin is influenced by not only the external magnetic field but also the nuclear magnetic field $\vec{B}_n$ due to the presence of $\vec{\mu}_n$. Therefore, the evolution of $\vec{\mu}_e$ in the IR chamber results from the combined effects of $\vec{B}_n$ and a quadrupole magnetic field $\vec{B}_q$, which is generated by combining the magnetic field from the wire and a vertical remnant fringe magnetic field $\vec{B}_r$ [6]. In SG apparatuses with a strong external magnetic field $\vec{B}_0$, the magnetic moments of the electron and nucleus precess in opposite directions at different speeds. As a result of induction in the electron–nuclear interaction, $\vec{\mu}_e$ is repelled by $\vec{\mu}_n$ to either align or anti-align with $\vec{B}_0$, in other words, collapses to the eigenstates. In contrast to precession, collapse can be regarded as the secondary motion of the electron magnetic moment. The electron spin collapses much faster than the nuclear spin due to the slower precession of $\vec{\mu}_n$, so we neglect the collapse of the co-quantum.

The co-quantum guides the collapse of the principal quantum according to their relative orientations at the time of measurement, yielding the CQD branching condition [12], [13]. CQD has been shown to statistically recover the quantum mechanical wave function from the continuous angular distribution of the co-quantum that is assumed to be isotropic for atoms immediately out of the oven (see Figure 1) [12], [13]. However, the angular distribution of the co-quantum is altered by the collapse of the principal quantum. Selection of one branch of an SG apparatus automatically selects the portion of the co-quanta that could guide the principal quanta to the specific eigenstate (see Results). Therefore, neither branch of the SG output has an isotropic distribution of co-quanta, which has not been considered in previous theories for multi-stage SG experiments [6], [7].

Here, we present a numerical simulation using the Schrödinger equation based on the torque-averaged magnetic field $\vec{B}_n$ of the co-quantum with a continuous angular distribution, along with the branching condition, to yield a CQD model for the flip of electron spin in the Frisch–Segrè experiment. Our numerical simulation agrees with the Frisch–Segrè experimental observation and confirms our previous closed-form analytical solution of CQD [12], [13], which builds on the insightful analysis by Majorana [6].

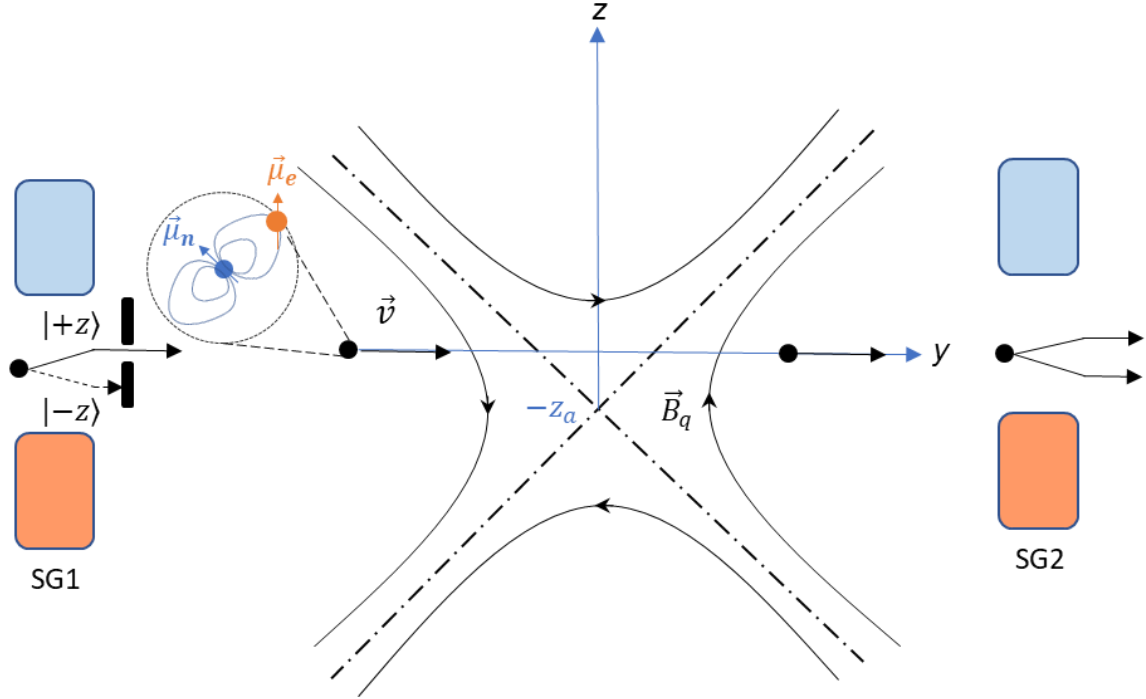

**Figure 1**. Illustration of the multi-stage SG experiment conducted by Frisch and Segrè. Potassium atoms from an oven are sent to the first SG apparatus (SG1). Conceptually, a slit is used to select the up branch of SG1; although this slit was physically placed after the quadrupole field in the experiment, we relocate it to simplify the analysis. $|+z\rangle$ and $|-z\rangle$ denote the up state and down state of $\vec{\mu}_e$, respectively. Atoms travel along the $y$-axis through a quadrupole magnetic field $\vec{B}_q$. The center (null point) of the quadrupole field is at $\vec{r}_{\text{null}} = (0,0,-z_a)$, where $z_a = 1.05 \times 10^{-4}$ m. Following the spin rotation by the quadrupole field, the second SG apparatus (SG2) measures the fraction of spin slip of the electron magnetic moment. $\vec{\mu}_e$, electric magnetic moment. $\vec{\mu}_n$, nuclear magnetic moment.

## Method

The Frisch–Segrè experiment consists of two SG apparatuses separated by an IR chamber, which rotates $\vec{\mu}_e$ using an approximate quadrupole magnetic field, as shown in Figure 1. Potassium atoms thermally effused from an oven enter the first SG apparatus, where they are split into up and down branches. The CQD branching condition states that if $\vec{\mu}_e$ is angularly closer (farther) to the $+z$-axis than $\vec{\mu}_n$, $\vec{\mu}_e$ is repelled by $\vec{\mu}_n$ to collapse parallel (antiparallel) to the $+z$-axis. Therefore, the pre-collapse state of $\vec{\mu}_e$ with a given $\vec{\mu}_n$ determines the measurement outcome according to the relative polar orientations of $\vec{\mu}_e$ and $\vec{\mu}_n$:

$$|\mu_e \copyright \mu_n\rangle = \frac{1 - \text{sgn}(\theta_e - \theta_n)}{2}|+z\rangle + \frac{1 + \text{sgn}(\theta_e - \theta_n)}{2}|-z\rangle, \tag{1}$$

where the prefix $\copyright$ indicates the co-quantum, sgn denotes the sign function, and $\theta_e$ and $\theta_n$ designate the polar angles of $\vec{\mu}_e$ and $\vec{\mu}_n$ in spherical coordinates in $\mathbb{R}^3$, respectively. We use $|+z\rangle$ and $|-z\rangle$ to denote the states where $\vec{\mu}_e$ is aligned with $+z$ or $-z$, respectively. Upon selecting the $|+z\rangle$ branch from SG1, the associated co-quanta follow an anisotropic probability density function [12]:

$$p_n(\theta_n, \phi_n) = \frac{1 - \cos\theta_n}{4\pi}. \tag{2}$$

At the entrance of the IR chamber, the initial state of $\vec{\mu}_e$ is $|+z\rangle$ with $(\theta_{e0}, \phi_{e0}) = (0,0)$, where $\theta_{e0}$ and $\phi_{e0}$ are the initial polar and azimuthal angles of $\vec{\mu}_e$ in spherical coordinates in $\mathbb{R}^3$. In contrast to conventional quantum mechanical descriptions [7], [8], CQD assumes that the angular distribution of the co-quanta is continuous. The initial polar and azimuthal angles $(\theta_{n0}, \phi_{n0})$ of $\vec{\mu}_n$ are sampled from Eq. (2) through the Monte Carlo method as follows:

$$\theta_{n0} = 2\arcsin\left(\zeta_1^{\frac{1}{4}}\right), \qquad \phi_{n0} = 2\pi\zeta_2, \tag{3}$$

where $\zeta_1$ and $\zeta_2$ are uniformly distributed random numbers between 0 and 1.

The IR chamber consists of a current-carrying wire along the $x$-axis and a remnant field $\vec{B}_r$ along the $z$-axis with a magnitude of $0.42 \times 10^{-4}$ T. The magnetic field generated by the wire and the remnant field cancel at $\vec{r}_{\text{null}} = (0,0,-z_a)$ of the coordinate system in Figure 1, forming an approximate quadrupole field [16]. The field gradient is found from the wire current $I$ and the remnant field to be [6]

$$G = \frac{2\pi}{\mu_0 I} B_r^2. \tag{4}$$

Let $t$ be the flight time of an atom traveling with velocity $v$ (800 m/s) and $z_a$ the vertical distance between the selected atomic beam and the wire ($1.05 \times 10^{-4}$ m). We set the time $t = 0$ when the atom is at the origin of the coordinate. The quadrupole field $\vec{B}_q$ at $\vec{r} = (0, vt, 0)$ is given by $\vec{B}_q = (0, Gz_a, Gvt)$.

In addition to the external field, the torque-averaged nuclear magnetic field given below also acts on the electron [12]:

$$\vec{B}_n = \frac{5\mu_0}{16\pi R^3} \vec{\mu}_n, \tag{5}$$

where $\mu_0$ is the vacuum permeability, $R = 2.75 \times 10^{-10}$ m is the van der Waals radius of the potassium atom [17], and $\mu_n = 1.977 \times 10^{-27}$ J/T is the magnitude of the nuclear magnetic moment. The total magnetic field experienced by $\vec{\mu}_e$ is $\vec{B} = \vec{B}_q + \vec{B}_n = (B_x, B_y, B_z)$:

$$B_x = B_n \sin\theta_n \cos\phi_n, \tag{6}$$

$$B_y = Gz_a + B_n \sin\theta_n \sin\phi_n, \tag{7}$$

$$B_z = Gvt + B_n \cos\theta_n. \tag{8}$$

The final orientation of $\vec{\mu}_e$ at the output of the IR chamber is numerically calculated through the Schrödinger equation with a modified Hamiltonian to account for the contribution of $\vec{\mu}_n$. Because we have shown that the Schrödinger equation for electron spin can be derived from the Bloch equation [13], the Schrödinger equation is used here as a mathematical tool; a direct numerical solution to the Bloch equation is reported separately [16].

The evolution of the state of the electron magnetic moment $|\mu_e\rangle = \begin{pmatrix} c_1 \\ c_2 \end{pmatrix} = \begin{pmatrix} \cos(\theta_e/2) \\ \sin(\theta_e/2)\exp(i\phi_e) \end{pmatrix}$ is governed by the Schrödinger equation:

$$i\hbar \frac{d}{dt} |\mu_e\rangle = H |\mu_e\rangle, \tag{9}$$

where $c_1$ and $c_2$ are the probability amplitudes of the $|+z\rangle$ and $|-z\rangle$ states, and the Hamiltonian is

$$H = -\frac{1}{2}\hbar\gamma_e \vec{B} \cdot \vec{\sigma}.$$  (10)

Substituting the Pauli vector $\vec{\sigma}$ yields

$$H = -\frac{1}{2}\hbar\gamma_e \begin{pmatrix} B_z & B_x - iB_y \\ B_x + iB_y & -B_z \end{pmatrix},$$  (11)

where $\gamma_e = -1.761 \times 10^{11}$ rad $\cdot$ Hz/T is the gyromagnetic ratio of the electron. To simplify Eq. (9), we replace the time $t$ with dimensionless time $\tau$:

$$\tau = \frac{1}{2}\sqrt{\gamma_e G v} \cdot t + \frac{1}{2}\sqrt{\frac{\gamma_e}{G v}} B_n \cos\theta_n .$$  (12)

We assume that $\theta_n \approx \theta_{n0}$ because the collapse of $\vec{\mu}_n$ is too slow to occur during the flight time. The azimuthal angle evolves as $\phi_n = w_n\tau + \phi_{n0}$, where the dimensionless Larmor frequency $w_n$ is given by

$$w_n = 2\frac{\gamma_n B_e}{\sqrt{\gamma_e G v}}.$$  (13)

Here, $B_e$ is the magnitude of the torque-averaged electron magnetic field on the nucleus given below [12]:

$$\vec{B}_e = \frac{5\mu_0}{16\pi R^3}\vec{\mu}_e$$  (14)

with $\mu_e = 9.285 \times 10^{-24}$ J/T.

To suppress high-frequency oscillations, Majorana defined the following transformation of variables [6]:

$$\begin{pmatrix} c_1 \\ c_2 \end{pmatrix} = \begin{pmatrix} e^{-i\tau^2}f \\ e^{+i\tau^2}g \end{pmatrix}.$$  (15)

This transformation not only simplifies the mathematical derivation but also accelerates our numerical simulation substantially.

Substituting Eqs. (11–15) into Eq. (9) yields

$$\frac{d^2f}{d\tau^2} - 4i\left[\tau - \frac{\sqrt{k_1}w_n}{4\left(\sqrt{k_1} - i\sqrt{k_0}e^{i\phi_n}\right)}\right]\frac{df}{d\tau} + \left(k_0 + k_1 + 2\sqrt{k_0 k_1}\sin\phi_n\right)f = 0, \tag{16}$$

where adiabaticity parameters $k_0$ and $k_1$ are defined as

$$k_0 = \frac{z_a}{v}\gamma_e G, \tag{17}$$

$$k_1 = \gamma_e \frac{(B_n \sin\theta_{n0})^2}{Gv}. \tag{18}$$

The initial conditions of Eq. (16) are $f(-\infty) = 1$ and $df(-\infty)/d\tau = 0$ [6]. For each sampled $(\theta_{n0}, \phi_{n0})$, the numerical solution is conducted over a dimensionless time range $-30 < \tau \leq 60$. Since the variation of $f$ is negligible when the atom is far before the null point, we consider the initial conditions as $f(-30) = 1$ and $df(-30)/d\tau = 0$ in the numerical simulation.

Majorana reasoned that because the $z$-component of the magnetic field is reversed along the flight path, the roles of $f$ and $g$ in the quantification of spin flip are reversed [6]; the justification that we found is the initial adiabatic flip when the atom passes above the wire [12], [16]. Therefore, we compute the final polar angle of $\vec{\mu}_e$ using the final value of $f$ through $|f| = \sin(\theta_{e,f}/2)$ [10], yielding $\theta_{e,f} = 2\arcsin|f|$. Substituting the final polar angles $\theta_e$ and $\theta_n$ into Eq. (1) predicts the collapsed state measured by SG2. According to the branching condition, the final orientation of $\vec{\mu}_e$ is

$$\theta_{e,D} = \begin{cases} 0 & \text{if } \theta_{e,f} < \theta_{n,0} \\ \pi & \text{if } \theta_{e,f} > \theta_{n,0} \end{cases}. \tag{19}$$

We sample $N = 2 \times 10^4$ sets of $(\theta_{n0}, \phi_{n0})$ and statistically calculate the fraction of $\vec{\mu}_e$ that collapse to $|{+}z\rangle$. The fraction of spin flip is given by

$$W_{\text{num}} = \frac{1}{N}\sum_{j=1}^{N}[\theta_{e,D}^j = 0]. \tag{20}$$

The Iverson bracket takes on 1 when the statement inside the bracket is true and 0 otherwise. Our source code written in Mathematica is provided in Appendix: Source code in Mathematica and online [18].

## Results

Figure 2 demonstrates the solutions of $|f(\tau)|$ versus $\tau$ with different wire currents for an example pair of $(\theta_{n0}, \phi_{n0})$. The electron magnetic moment rotates its polar angle mostly near the quadrupole null point, and $|f(\tau)|$ oscillates with a damping amplitude thereafter. The final value of $|f(+\infty)|$ is estimated by averaging $|f(\tau)|$ in the range $52 < \tau \le 60$.

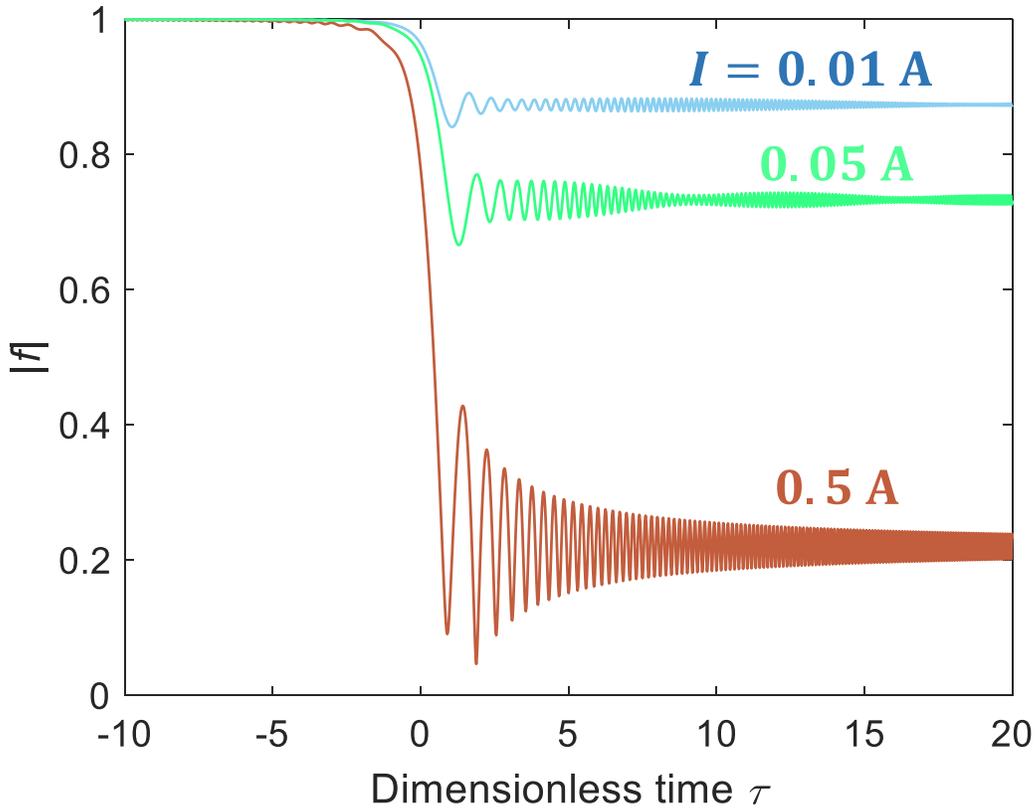

**Figure 2**. Examples of $|f(\tau)|$ versus $\tau$ with different wire currents $I$ but the same initial orientation $(\theta_{n0}, \phi_{n0}) = (6\pi/7, 0)$.

Two probability density functions for $\vec{\mu}_n$ are considered: isotropic ($p_{n,\text{isotropic}} = 1/4\pi$) and anisotropic ($p_{n,\text{anisotropic}} = (1 - \cos\theta_{n0})/4\pi$). These results are compared with the Frisch–Segrè observation in Figure 3. The isotropic distribution results in a negative coefficient of determination $R^2_{\text{isotropic}} = -0.26$, indicating poor agreement, whereas the anisotropic distribution shows a high coefficient of determination $R_{\text{anisotropic}}^2 = 0.95$. Similarly, the coefficient of determination for the closed-form analytical solution [12] is $R_{\text{analytical}}^2 = 0.96$.

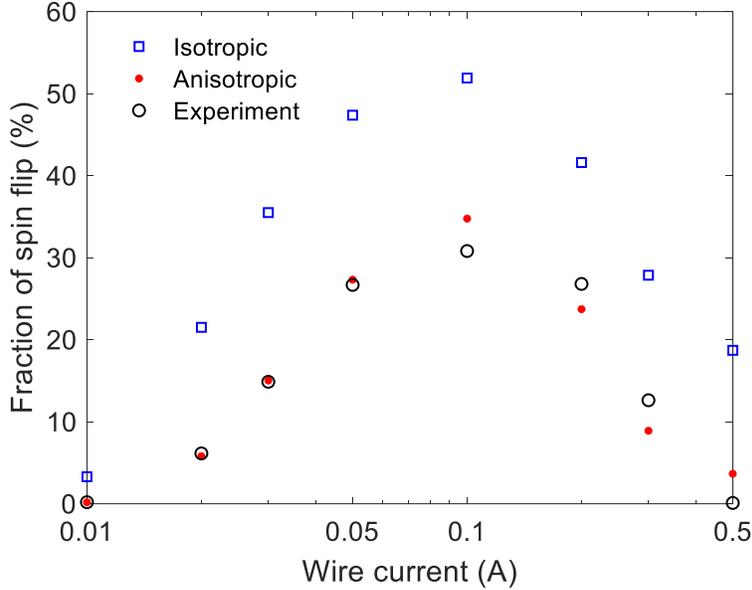

**Figure 3**. Simulated and experimental results for the fraction of spin flip in the Frisch–Segrè experiment. Isotropic and anisotropic distributions of the nuclear magnetic moment yield the coefficients of determination of $R_{\text{isotropic}}^2 = -0.26$ and $R_{\text{anisotropic}}^2 = 0.95$, respectively. The number of simulated atoms is $2 \times 10^4$, leading to error bars smaller than the size of the symbols.

## Discussion and conclusion

The numerical model incorporates the CQD concept into the Schrödinger equation [12]. The co-quantum $\vec{\mu}_n$ remains continuous upon the collapse of $\vec{\mu}_e$, rather than becoming quantized. In this simulation, we approximate the precession of $\vec{\mu}_n$ with a constant Larmor frequency. Solving the Bloch equation with a variable Larmor frequency, which however is numerically stiff, has led to similar agreement with the experimental observation [16]. By contrast, Rabi's quantum mechanical formula based on a quantized isotropic distribution for the nuclear magnetic moment does not fully agree with the experimental observation [12]. As shown in another paper, the standard quantum mechanical treatment using the von Neuman equation does not yet match the experimental observation [19].

In conclusion, we demonstrate a numerical model for the multi-stage Stern–Gerlach experiment by Frisch and Segrè based on CQD. The collapse of electron magnetic moment in SG1 according to the CQD branching condition leads to the redistribution of the co-quantum. We apply the Monte Carlo method to sample the co-quantum. With a Hamiltonian modified with the torque-averaged magnetic field from the co-quantum, we use the Schrödinger equation to model the evolution of the electron magnetic moment inside the IR chamber. Then, the branching condition is applied to SG2 to quantify the fraction of spin flip. Our numerical model closely predicts the observation of the Frisch–Segrè experiment with no free parameters.

## Data availability

All data used in this study are available from the corresponding author upon reasonable request.

## Code availability

The custom source code written in Mathematica for this study is available in Appendix and online [18].

## Acknowledgments

This project has been made possible in part by grant number 2020-225832 from the Chan Zuckerberg Initiative DAF, an advised fund of the Silicon Valley Community Foundation.

## Author contributions

Z.H., K.T., and D.G. developed the simulations and analyzed the data. S.K. helped with the validation of the simulations. L.V.W. conceived the concept and developed the theory as well as closely supervised the project. All authors contributed to writing the manuscript.

## Competing interests

The authors declare no competing interests.

# Appendix: Source code in Mathematica

```
IwireList = List[0.01, 0.02, 0.03, 0.05, 0.1, 0.2, 0.3, 0.5];              (*current*)
v = 800.;                                                                  (*velocity*)
m0 = N[4.*Pi*1*^-7, 16];                                        (*vacuum permeability*)
R = 2.75*^-10;                                                        (*atom radium*)
mun = 1.97723*^-27;                                         (*nuclear magnetic moment*)
mue = 9.28*^-24;                                           (*electron magnetic moment*)
za = 1.05*^-4;                                    (*distance between wire and atom beam*)
Be = N[5 m0 mue/(16 Pi*R^3), 16];                           (*magnetic field of mu_e*)
Bn = N[5 m0 mun/(16 Pi*R^3), 16];                           (*magnetic field of mu_n*)
gammae = 1.76085963*^11;                            (*gyromagnetic ratio of electron*)
gamman = 1.25*^7;                                      (*gyromagnetic ratio of nuclei*)
Br = 0.42*1.*^-4;                                          (*remnant magnetic field*)
tN = 2051;                    (*2051 time sequence; 20000 for plotting the smooth sequence*)
pCQD4List = ConstantArray[0, 8];                          (*accumulated flipped spins*)
S = 2000;                                                      (*number of samplings*)
numberOfNP = 1; n = 1;                                      (*usually they are 1*)
taumin = -30;                                          (*dimensionless time start*)
taumax = 60;                                              (*dimensionless time end*)
tauList = taumin + Range[tN]/tN*(taumax - taumin);              (*time sequence*)
SEListManual = ConstantArray[0, {S, 8}];                  (*initialize the spin flip*)
fList = ConstantArray[0, {tN - 1, 8}];              (*initialize the time sequence*)
Do[
sampleOrder = s;                                      (*sample order, s from 1 to 2000*)
w2plotMatrix = ConstantArray[0, {numberOfNP, 8}]; (*initial spin flip for one sampled atom*)
length = (Range[numberOfNP])/numberOfNP;

Do[len = length[[m]];
Do[Iwire = IwireList[[j]];
PflipCQD4 = 0;
Do[thn = 2*ArcSin[(RandomReal[])^(1/4)];
(*sample polar angle of the nuclear magnetic moment, the example is the anisotropic
distribution*)
phia = RandomReal[]*2 Pi;                                  (*azimuthal angle mu_n*)
G = 2 Pi (Br)^2/(m0* Iwire);                          (*G quadrupole gradient*)
k0 = Abs[gammae]*za^2/v* G;                                        (*k0*)
k1 = Abs[gammae]*(Bn Sin[thn])^2/(G v);                            (*k1*)
\[Alpha] = 1/2*Sqrt[Abs[gammae ] G v];        (*ratio between time and dimensionless time*)
tNP = Bn Cos[thn]/(G v);                      (*time when atom reaches the null point*)
fInitial = 1;                                          (*initial condition*)
DfInitial = 0;                                          (*initial condition *)
Do[
tau1 = tauList[[q]];                                      (*time step q*)
tau2 = tauList[[q + 1]];                                  (*time step q+1*)
```

```
\[Omega]n =   Abs[gamman]*Be;                          (*precession frequency of mu_n*)
wn = \[Omega]n /\[Alpha];                          (*dimensionless precession frequency*)
phin = wn*tau + phia;                                    (*azimuthal angle of mu_n*)
sparam1 = NDSolve[{
f1''[tau] == ((Sqrt[k1] wn / (I Sqrt[k1] + Sqrt[k0] Exp[I (phin)])) +
4 I (tau)) f1'[tau] - (k1 + k0 + 2 Sqrt[k0 k1] Sin[phin]) f1[tau],
f1[tau1] == fInitial, f1'[tau1] == DfInitial},
  f1,
  {tau, tau1, tau2},
Method -> Automatic,
PrecisionGoal -> 10,
AccuracyGoal -> 10, InterpolationOrder -> All,
WorkingPrecision -> MachinePrecision, MaxSteps -> Infinity];        (*Schrodinger equation*)
fList[[q, j]] = Max[Abs@f1[tau1] /. sparam1]; (*time sequence at q's step for j's current*)
fInitial = Max[f1[tau2] /. sparam1];
DfInitial = Max[f1'[tau1] /. sparam1], {q, 1, tN - 1}];
getFfinal =  Mean[fList[[tN - 187 ;; tN - 1, j]]];                        (*average f*)
thetaE = Re[2 ArcSin[getFfinal]];                     (*calculate the polar angle of mu_n*)
getCQDflipover = Evaluate[(Sign[thetaE - thn] + 1)/2];       (*branching condition Eq. (1)*)
PflipCQD = getCQDflipover + PflipCQD4, {i, 1, n}];
pCQD4List[[j]] = PflipCQD4/n*100;                          (*probability in %*)
, {j, 1, 8}];
w2plotMatrix[[m]] = pCQD4List, {m, 1, numberOfNP}];
fileOrder = 1;                                            (*usually equals to 1*)
SEListManual[[sampleOrder + 1, All]] = w2plotMatrix[[fileOrder]];
, {s, 0, S - 1} ]

(*Plot*)
SEresults =
ListLogLinearPlot[Transpose@{IwireList, (Mean[SEListManual])},
PlotLegends -> {"SE Results"}, PlotStyle -> Blue,
PlotMarkers -> "O", Joined -> True];
```